\documentclass[%
 reprint,
superscriptaddress,
amsmath,amssymb,
aps
]{revtex4-2}

\usepackage{graphicx}% Include figure files
\usepackage{dcolumn}% Align table columns on decimal point
\usepackage{bm}% bold math
\usepackage[version=4]{mhchem}
\usepackage{color}

\begin{document}

%\preprint{APS/123-QED}
%TC:ignore 
\title[Nanosecond laser annealing: impact on superconducting silicon on insulator epilayers
]{Nanosecond Laser Annealing: impact on superconducting Silicon on Insulator epilayers}

\author{Y. Baron}
\affiliation{Uni. Paris-Saclay, CNRS, Centre de Nanosciences et de Nanotechnologies, 91120, Palaiseau, France}
\author{J. L. L\'ab\'ar}
\affiliation{Thin Film Physics Laboratory, Institute of Technical Physics and Materials Science, Centre of Energy Research, Konkoly Thege M. u. 29-33, H-1121 Budapest, Hungary}
\author{S. Lequien}
\affiliation{Uni. Grenoble Alpes, CEA,IRIG-MEM, 38000 Grenoble, France}
\author{B. Pécz}
\affiliation{Thin Film Physics Laboratory, Institute of Technical Physics and Materials Science, Centre of Energy Research, Konkoly Thege M. u. 29-33, H-1121 Budapest, Hungary}
\author{R. Daubriac}
\affiliation{Univ. Grenoble Alpes, CEA,LETI, Minatec Campus, 38000 Grenoble, France}
\author{ S. Kerdilès}
\affiliation{Univ. Grenoble Alpes, CEA,LETI, Minatec Campus, 38000 Grenoble, France}
\author{P. Acosta ALba}
\affiliation{Univ. Grenoble Alpes, CEA,LETI, Minatec Campus, 38000 Grenoble, France}
\author{C. Marcenat}
\affiliation{Univ. Grenoble Alpes, CEA, Grenoble INP,IRIG-PHELIQS, 38000 Grenoble, France}
\author{D. Débarre}
\affiliation{Uni. Paris-Saclay, CNRS, Centre de Nanosciences et de Nanotechnologies, 91120, Palaiseau, France}
\author{F. Lefloch}
\email[ Corresponding author: ]{francois.lefloch@cea.fr}
\affiliation{Univ. Grenoble Alpes, CEA, Grenoble INP,IRIG-PHELIQS, 38000 Grenoble, France}
\author{F. Chiodi}
\email[ Corresponding author: ]{francesca.chiodi@c2n.upsaclay.fr}
\affiliation{Uni. Paris-Saclay, CNRS, Centre de Nanosciences et de Nanotechnologies, 91120, Palaiseau, France}

\date\today

\begin{abstract}
	\noindent We present superconducting monocrystalline Silicon On Insulator thin 33 nm epilayers. They are obtained by nanosecond laser annealing under ultra-high vacuum on 300 mm wafers heavily pre-implantated with boron ($2.5\times \,10^{16}\, at/cm^2$, 3 keV). Superconductivity is discussed in relation to the structural, electrical and material properties, a step towards the integration of ultra-doped superconducting Si at large scale. In particular, we highlight the effect of the nanosecond laser annealing energy and the impact of multiple laser anneals. Increasing the energy leads to a linear increase of the layer thickness, and to the increase of the superconducting critical temperature $T_c$ from zero ($<35\, mK$) to $0.5\,K$. This value is comparable to superconducting Si layers realised by Gas Immersion Laser Doping where the dopants are incorporated without introducing the deep defects associated to implantation. Superconductivity only appears when the annealed depth is larger than the initial amorphous layer induced by the boron implantation. The number of subsequent anneals results in a more homogeneous doping with reduced amount of structural defects and increased conductivity. The quantitative analysis of $T_c$ concludes on a superconducting/ non superconducting bilayer, with an extremely low resistance interface. This highlights the possibility to couple efficiently superconducting Si to Si channels.   
\end{abstract}

\maketitle
%TC:endignore

\section*{Introduction}
In the context of solid-state based quantum engineering, material science remains a very active field of research. Recent reviews \cite{Goh2022, Polini2022,NatureReviewsMaterials2021, Leon2021} point out the importance of controlling both the quality of materials, to prevent/reduce quantum decoherence \cite{Siddiqi2021, Place2021}, and the reproducibility, in the prospect of scaling-up quantum technology towards a very large number of qubits. 
In this quest of large scale integration, silicon and germanium can be seen as the short-term most promising materials \cite{Vinet2021,Scappucci2021}. However, they are now restricted to spin qubits, where the rather low yield in the quantum properties has limited the demonstration of coupled qubits to a very small number \cite{Philips2022}, as compared to superconducting transmons qubits where operations with few tens of qubits have been recently demonstrated \cite{Arute2019,Wu2021}. The possibility to fabricate superconducting qubits with silicon would give both the advantage of a mature technology and of superconductivity. In principle, this could be done with superconducting silicon, obtained by combining heavily boron doping and nanosecond laser annealing \cite{Bustarret2006, Marcenat2010, Chiodi2014, Daubriac2021,Dumas2023}. But implementation of superconducting Si:B into quantum circuits is still at its early age \cite{Duvauchelle2015, Chiodi2017} and a better understanding of the superconducting properties is still required, especially with the constraints of the use of compatible large scale integration tools. 
In the present study, we investigate the superconducting properties of boron doped superconducting silicon epilayers obtained on 33 nm thick SOI (Silicon On Insulator) 300 mm wafers after pre-implantation of boron dopants followed by nanosecond laser annealing. Our results demonstrate a continuous increase of the superconducting critical temperature $T_c$ as a function of melted depth tuned by the laser energy, and the weak impact of implantation-induced defects. A maximum $T_c$ of about $0.5\,K$ is reached when the entire 33 nm thick silicon layer is melted, producing a crystalline structure, while no superconductivity is observed when the annealing only affects the surface amorphous layer induced by the strong dopant implantation. We emphasize the role of identical multiple laser shots by comparing series of sample having sustained 1 or 5 laser shots annealings, highlighting the decrease of defects and better homogeneity with an increased number of shots. The $T_c$ variation is well described by a two-layers model (one being superconducting and the other not) connected through a high transparency interface. This points out the possibility to optimally couple the superconducting layer to silicon channels through low resistance interfaces.   

\subsection*{Nanosecond annealing of B implanted SOI}

	\begin{figure}[h]
		\centering
		\includegraphics[width=0.88\columnwidth]{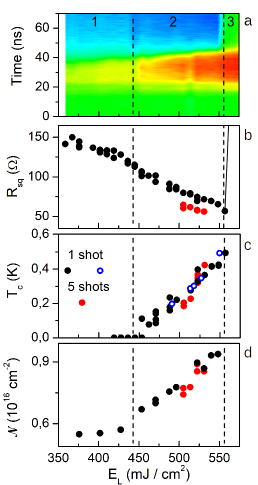}
		\caption{\label{fig:Rs_Tc_Dose_vs_Elaser} Combined plot of Time Resolved Reflectometry $TRR$ (a), room temperature square resistance $R_{sq}$ (b), superconducting critical temperature $T_c$ (c) and measured active carriers surface density ($\mathcal{N}$) (d) as a function of the laser energy density $E_L$. The dotted line at $E_L =445 \, mJ/cm^2$ indicates the cross-over from a polycrystalline boron doped layer (regime 1) to a crystalline epilayer (regime 2). The dotted line at $E_L = 560 \, mJ/cm^2$, where   $R_{sq}$ diverges, corresponds to the full melt of the initial silicon on insulator layer as the annealing touches the amorphous $SiO_2$ BOX (regime 3). The $TRR$ color plot scale corresponds to the measured TRR amplitude from 0.75$\,\mathcal{R}_{Si}$ (dark blue) and 2.26$\,\mathcal{R}_{Si}$ (red), and is a signature of the layer structure: amorphous (green), polycristalline (light blue), monocrystalline (dark blue), liquid (red). The laser pulse starts at t=10 ns.   
  }
	\end{figure}	

The first step of the fabrication of superconducting SOI layers is the implantation of a very high boron dose ($2.5\times10^{16} \,$cm$^{-2}$) at $3\,$keV on a non-intentionally doped $300\,$mm 
wafer. The initial SOI wafer has a silicon layer $33\,$nm $\pm 1\,$ nm thick on top of a SiO$_2$ buried oxide layer (BOX) of $20\,$nm, as in \cite{Daubriac2021}.  After the implantation, the SOI layer is composed of an amorphous layer (a-Si) of $15 \pm 1\, $nm on top of the remaining silicon crystalline layer (c-Si) (see Fig.4a in \cite{Daubriac2021}). CTRIM simulations show that the dopants concentration in the underneath c-Si layer remains significant (above $1\,at.\%$ percent) \cite{Posselt1992}.  \\
The nanosecond laser annealing process is performed with an excimer XeCl laser of pulse duration $25\,$ns under UHV (Ultra High Vacuum) conditions ($P=10^{-9}\,$mbar) with energy density at the sample level $E_L = 300$ to $600\,mJ/cm^2$ ($E_L = 600\,mJ/cm^2$ corresponding to bulk Si melting threshold). The effect of the laser pulse is to melt the top of the implanted silicon layer, over a $2$mm$\, \times\, 2$mm surface with laser energy homogeneity of $1.2 \%$, during $\approx 15$ to $25\,$ns. This induces an extremely fast re-crystallisation, activating the dopants up to a saturation concentration of $n_{sat} = 3\times10^{21} \,$cm$^{-3}$ (6 at.$\%$),  an order of magnitude above the solubility limit $n_{sol}\sim 4\times10^{20}\,$cm$^{-3}$ \cite{Desvignes2023}. The thickness of the melted layer linearly depends on the laser energy at the sample level. The Time Resolved Reflectometry (TRR) of a red ($\lambda = 675\,$nm) laser is recorded in-situ during the nanosecond laser annealing to follow the melting-solidification process, and its value is compared to bare, undoped, Si reflectivity $\mathcal{R}_{Si}$ (Fig.\ref{fig:Rs_Tc_Dose_vs_Elaser}a).\\ 
For the present study, two series of laser spots have been generated. For the first series, each spot has been patterned after the laser annealing, to define Hall cross structures with Ti/Au contact pads allowing precise measurement of the square and Hall resistances. The second series of spots remained untreated, for X-Rays Diffraction (XRD) analysis. For both series, we have measured the resistance as a function of temperature down to $\approx 35\, mK$ to extract the superconducting transition temperature $T_c$ as a function of the laser energy $E_L$. We emphasize the precise control on $E_L$, and the impact of the sharp, flat, few nanometer thick interface at the bottom of the annealed layer, which allow to fine-tune with $E_L$ the depth in the 33 nm deep SOI with 1-2 nm precision. The overall results for the time resolved reflectivity $TRR$, the square resistance $R_{sq}$, the superconducting transition temperature $T_c$ and the active dopants dose $\mathcal{N}$  measured by Hall effect are shown in Fig.\ref{fig:Rs_Tc_Dose_vs_Elaser} as a function of the laser energy density 
$E_L$. \\
We can identify three regimes. The first regime (regime 1) is for $E_L < 445\,mJ/cm^2$ where no superconductivity is observed. The second regime (regime 2) applies for $445\,mJ/cm^2 < E_L < 560\,mJ/cm^2$ with an almost linear increase of $T_c$ up to a maximum value of $0.5\,K$. For $E_L > 560\,mJ/cm^2$ (regime 3), the entire silicon layer is full-melted, the resistance diverges and no superconductivity is observed. 
We observe important differences to the results of \cite{Daubriac2021}, where the annealing was performed on the $same$ implanted layers under $N_2$ with a similar XeCl excimer laser but of longer laser pulse duration (160 ns instead of 25 ns). In the present work, we achieve a $T_c$ tunable up to 0.5 K, with a superconducting phase only observed in monocrystalline layers below the full melt threshold, as opposite to the constant $T_c \sim 0.18 \,K$ obtained only in a poly-crystalline layer above the full melt in \cite{Daubriac2021}. Our results are instead more similar to what was observed for the longer annealing of 160 ns with an increased ion implantation energy, from 3 keV to 4 keV, for which monocrystalline films were shown to be superconducting at $T_c=0.39\,$mK \cite{Dumas2023}. The recrystallisation dynamics seems then to be strongly dependent on the laser pulse duration, even for relative small changes (25 ns to 160 ns).
\subsection*{Explosive crystallisation, monocrystalline SiB, full-melt}\label{regimes}
Fig.\ref{fig:Rs_Tc_Dose_vs_Elaser} shows the drastically different behavior observed in the three regimes: the explosive crystallization regime (1), the monocrystalline regime (2), and the full melt regime (3).
In regime $1$ an explosive re-crystallisation of the doped amorphous Si (a-Si) takes place, followed by the partial melting of the resulting poly-crystalline Si (poly-Si). In this dynamic scenario \cite{Thomson1984}, even for weak laser energy, a thin layer of the initial top a-Si melts. The transition from solid to liquid occurs during the first nanoseconds of the laser pulse. This thin layer re-solidifies almost instantly into poly-Si and the latent heat released during this liquid to solid transformation is sufficient to progressively melt the a-Si underneath. The process stops when the entire initial a-Si is totally transformed into poly-Si, as the energy is not sufficient to melt crystalline Si. During the remaining time of the laser pulse, as the laser energy increases above the poly-Si melting threshold, the just-formed poly-Si is melted over a depth that depends linearly with the laser energy density. In regime $1$, the entire a-Si is therefore transformed into poly-Si with a final thickness independent on the laser energy, with part of the poly-Si having been remelted. This results in a variation of reflectivity from the initial amorphous state at $\mathcal{R}_{a}=$1.52$\,\mathcal{R}_{Si}$ to the polycrystalline state $\mathcal{R}_{poly}=1.05\,\mathcal{R}_{Si}$  (Fig.\ref{fig:Rs_Tc_Dose_vs_Elaser}a).    The square resistance varies only slightly with the laser energy density and the small decrease observed is due to the re-arrangement of poly-crystals and a slightly better activation as a consequence of the poly-Si melting (Fig.\ref{fig:Rs_Tc_Dose_vs_Elaser}b).   Correspondingly, the active dose ($\mathcal{N}$) measured by Hall effect is nearly constant and independent on $E_L$ (Fig.\ref{fig:Rs_Tc_Dose_vs_Elaser}c). From the laser energy density that characterizes regime 1, we estimate the thickness of the poly-Si layer to $\approx 18.5 \,nm$, close to the initial a-Si layer thickness \cite{Daubriac2021}.        
\newline

In regime $2$, the laser energy density is strong enough to melt the entire poly-Si created by explosive recrystallization from the a-Si layer, plus a thin part of the crystalline c-Si layer underneath. In that situation, the poly-Si is entirely melted and transformed into a boron doped monocrystalline silicon Si:B epilayer from the remaining weakly doped c-Si seed. This can be observed in the $TRR$ map as the reflectivity first peaks as a result of the explosive crystallisation (Fig.\ref{fig:Rs_Tc_Dose_vs_Elaser}a, yellow line at $t\sim 20\,$ns, $\mathcal{R}_{ec}=1.82\,\mathcal{R}_{Si}$) then increases to the melted phase (red, $\mathcal{R}_{melt}=2.2\,\mathcal{R}_{Si}$) to finally cool down to a monocrystal (blue, $\mathcal{R}_{mono}=0.95\,\mathcal{R}_{Si}$). The final thickness of Si:B layer on top of the c-Si depends linearly on the laser energy density (Fig.S\ref{fig:d_Elaser}). 
The active dose ($\mathcal{N}$) increases with $E_L$ (Fig.\ref{fig:Rs_Tc_Dose_vs_Elaser}d), due to the incorporation and activation of an increasing amount of implanted dopants in the Si:B layer. Note that the measured $\mathcal{N}$ is always lower than the initial implanted dose (at most 38$\%$), as the Hall effect only measures the active dose. From the $TRR$ maps identification of the full melt ($d_{melt} = d_{SOI}$) and the corresponding dose $\mathcal{N}_{FM}$), it is possible to extract the active concentration of the thickest monocrystalline layer, $\mathcal{N}_{FM}/d_{SOI} = 2.9\times 10^{21} cm^{-3}$. The active concentration for the thinnest monocrystalline layer is likewise calculated to $3.1\times 10^{21} cm^{-3}$. This suggests that the layers present a nearly constant active concentration $n_{sat}\approx 3\times 10^{21} cm^{-3}$, equal within $7\,\%$ for all the layers, as also confirmed by the nearly constant position of the Si:B XRD peak (Fig.S\ref{fig:XRD_spectra}). Similarly, a saturation of the active concentration $n_{sat}=2.8-3.1\times 10^{21} cm^{-3}$ was observed in bulk Si:B layers in the same 23 to 47 nm thickness range \cite{Desvignes2023}. Those layers, realized by Gas Immersion Laser Doping (GILD), employed nanosecond laser annealing with exactly the same laser, but with a BCl$_3$ gas precursor. The active concentration limit could thus be associated not to the dopant incorporation method, but to the maximum recrystallisation speed ($\sim 4 m/s$) \cite{Wood1984} induced by the 25 ns pulse duration. \\
As the Si:B layer becomes thicker, the square resistance decreases smoothly with $E_L$ (Fig.\ref{fig:Rs_Tc_Dose_vs_Elaser}b). One can notice that the square resistance after one laser shot is larger than after five shots. This can be understood as, on one side, the crystalline quality slightly improves upon repeated annealing, and, on the other, the homogeneity of the boron is increased as the dopants initially present in the top of the layer have more time to diffuse within the whole melted layer. 
In this regime $2$, the superconducting temperature transition grows from $T_C < 35\,mK$ ($35\,mK$ is the minimum transition temperature we could measure in our cryostat) to $0.5\,K$ just before reaching the full-melt regime $3$ (Fig.\ref{fig:Rs_Tc_Dose_vs_Elaser}c). Contrary to the $R_{sq}$ behavior, the $T_c$ difference between 1 and 5 shot(s) is less marked. \\

Finally, the regime $3$ is reached when the laser energy is large enough that the entire silicon layer melts (full melt threshold) and an amorphous Si:B layer builds up on the underneath amorphous SiO$_2$, recovering the initial amorphous reflectivity $R_{a}/R_{in}= 1.06$ (Fig.\ref{fig:Rs_Tc_Dose_vs_Elaser}a). As a result, the $R_{sq}$ diverges up to 100 k$\Omega$ and superconductivity is suppressed (Fig.\ref{fig:Rs_Tc_Dose_vs_Elaser}b,c).
\\

\subsection*{\label{sec:TcvsN} Superconductivity and number of laser pulses}
\begin{figure}[t]
\centering
\includegraphics[width=\columnwidth]{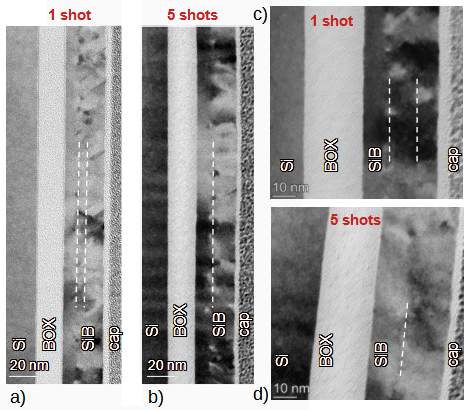}
\caption{\label{fig:TEM} TEM images of two samples realised with 1 (a,c) or 5 (b,d) laser pulses at $E_L$=522 mJ/cm$^2$ ($d_{melt}$=29 nm). From left to right are visible the Si substrate, the SiO$_2$ BOX, the B-implanted Si layer where the laser annealing took place, and a capping layer for TEM. The dotted lines mark where dislocations and stacking faults align, as a result of strain relaxation.}
\end{figure}

In order to better understand the impact of multiple laser shots on both the superconducting and the normal state, we complemented our transport measurements and X-Ray Diffraction data with Transmission Electronic Microscopy (TEM). We show in Fig.\ref{fig:TEM} the comparison between two samples in regime 2 having the same active dose, realised with the same laser energy $E_L$=522 mJ/cm$^2$ where nearly all the Si has been melted ($d_{melt}$=29 nm), with 1 or 5 laser shots.
The TEM images show from left to right the Si substrate, the SiO$_2$ BOX, the B-implanted Si layer where the laser annealing took place, and a capping layer introduced to protect the layer during TEM sample preparation. The interface between melted and unmelted SiB is difficult to discern in the contrast of this particular samples as the interface is only $\sim$2 nm away from the BOX, and as the doping increases gradually over a few nanometers from the bottom of the SiB layer \cite{Hallais2023}. 
The impact of performing a few laser shots as opposite to a single one is evident in the different sublayer structure within the laser annealed SiB, and in an increased disorder for 1 laser shot.  A larger amount of structural defects is present for a single laser annealing, with dislocations and stacking faults starting from two interfaces situated about 10 nm and 17-21 nm above the BOX, interfaces created between a fully strained layer on the bottom, a partially relaxed one in the middle with a gradually increasing deformation, and a top sublayer with a constant deformation (Fig.S\ref{fig:Strain}). In comparison, the sample realised with 5 laser shots is more homogeneous, with fewer defects starting 13 nm above the BOX, and lower deformation (Fig.S\ref{fig:Strain}). We can understand this difference as stemming from an incomplete B diffusion, from the higher implanted concentration at the top towards the bottom, during the short ($\sim 20 \,ns$) melting time for a single laser shot, as opposed to a more homogeneous dopant concentration for 5 laser shots allowing a longer diffusion time ($\sim 100 \,ns$). \\
These observations are in agreement with the electrical measurements: 1 shot samples show, systematically, a higher resistance (Fig.\ref{fig:Rs_Tc_Dose_vs_Elaser}) and a Residual Resistance Ratio RRR=$R_{300K}/R_{4K}$ closer to unity, indicating that the resistance at low temperature can be mostly attributed to impurities and crystallographic defects as opposed to thermal scattering. As an example, the two samples shown in Fig.\ref{fig:TEM} have RRR=1.1 and RRR=1.28 for 1 shot and 5 shots respectively. Similarly, XRD data show a better crystalline quality for the 5 shot samples, highlighted by a higher amplitude of the diffraction peak (Fig.S\ref{fig:XRD_spectra} and S\ref{fig:XRD_param}). \\
\begin{figure}[t]
		\centering
		\includegraphics[width=\columnwidth]{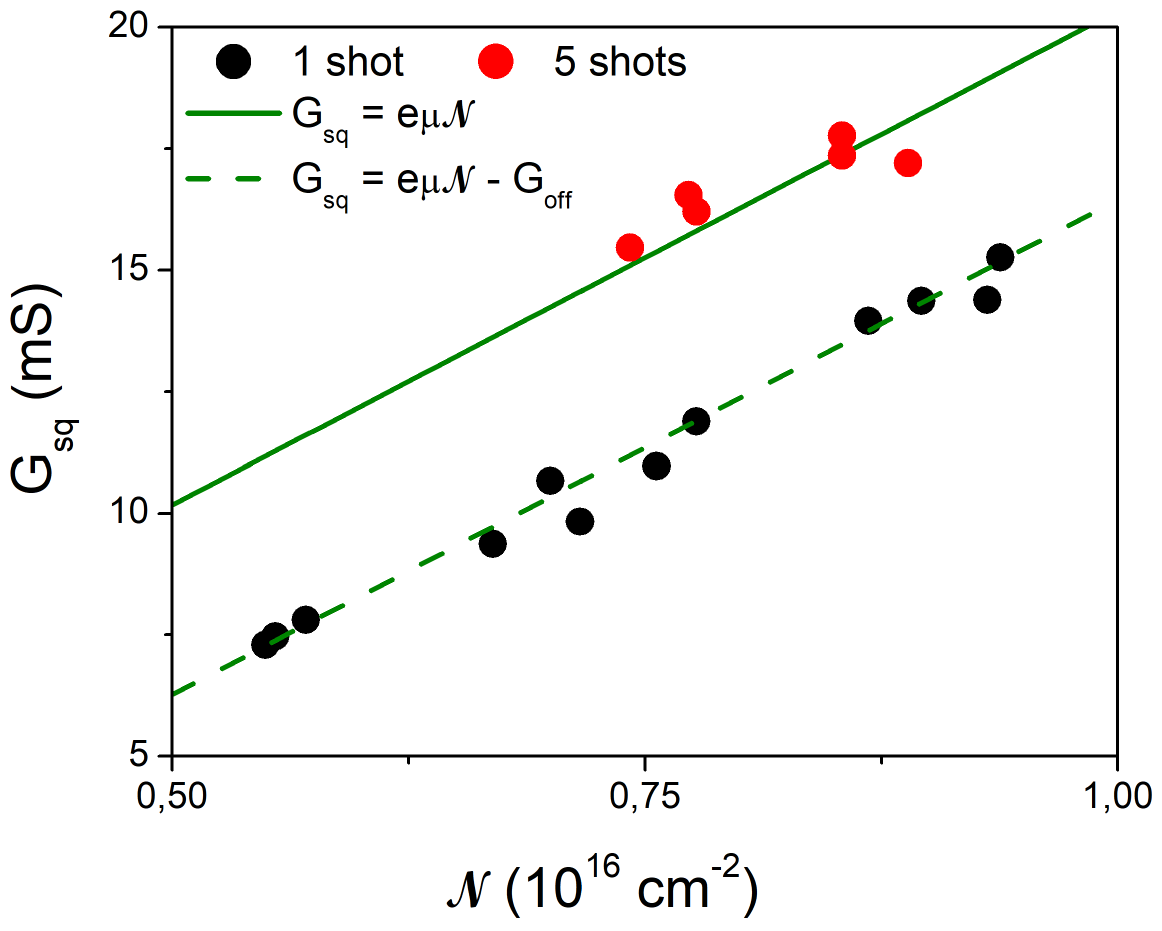}
		\caption{\label{fig:Gsq_vs_Dose} Square conductance (inverse of $R_{sq})$ as a function of the active dopants density $\mathcal{N}$ measured with Hall effect for both 1 and 5 shots series. The dotted line is a linear fit on the 1 shot series from which we can extract a mobility of $12.7\,cm^2V^{-1}s^{-1}$. The solid line has the exact same slope (corresponding to the same mobility) but passes though the origin of the graph at (0,0). The negative offset corresponds to an additional scattering mechanisms in the 1 shot series and is not present for the 5 shots series. }
	\end{figure}
In addition, we have plotted (Fig.\ref{fig:Gsq_vs_Dose}) the square conductance $G_{sq}$ ($1/R_{sq}$) as a function of the dose $\mathcal{N}$ for two series of 1 shot and 5 shots samples.
The two series show a linear $G_{sq} (\mathcal{N})$ dependence, which is consistent with the simple Drude formula $G_{sq}= e \mu \mathcal{N}$, where $e$ is the electronic charge and $\mu$ the carriers mobility. The mobility obtained by the linear fit is $\mu = 12.7\,cm^2 V^{-1}s^{-1}$, a value in agreement with past measurements for such doping levels \cite{Sy1985}.  However, this linear fit does not extrapolate to the x-axis origin for the 1 shot series (dotted line), whereas it does for the 5 shots series (solid line). The difference can be explained by the existence of a bilayer (or multilayer) structure for the 1 shot series, with one layer more disordered than the other. Indeed then $G_{sq}^{1shot}= e \mu \mathcal{N}_1 + e \mu' \mathcal{N}_2 < G_{sq}^{5shots}=e \mu \mathcal{N} $ with $\mu'<\mu$. Plotting $G_{sq}$ vs $\mathcal{N}=\mathcal{N}_1+\mathcal{N}_2$, gives $G_{sq}^{1shot}= e \mu (\mathcal{N}_1+\mathcal{N}_2) - e (\mu-\mu') \mathcal{N}_2$, with a negative offset such as the one observed in Fig.\ref{fig:Gsq_vs_Dose}. From the offset amplitude it is also possible to roughly estimate the thickness of the poorly conducting layer: taking the saturation active concentration in the whole annealed layer ($n\sim 3 \times 10^{21}\,$cm$^{-3}$) and assuming $\mu>>\mu' \sim 0$, from $\Delta G_{sq} = 3.9\,$mS = $e\mu\mathcal{N}_2 = e \mu n_2 d_2 $, we deduce $d_2 \sim 6.5 \,$nm.
For such rough estimation, this layer thickness is comparable, while smaller, to the 7 to 11 nm thickness of the central sublayer in 1 shot sample (Fig.\ref{fig:TEM}), where a high density of stacking faults and dislocations is observed, suggesting that the very disordered layer has a small (non-zero) $\mu'$, while the 5 shots samples have an homogeneous mobility. \\

\subsection*{\label{sec:modelling} Superconductivity and laser energy}

\begin{figure}[t]
		\centering
		\includegraphics[width=\columnwidth]{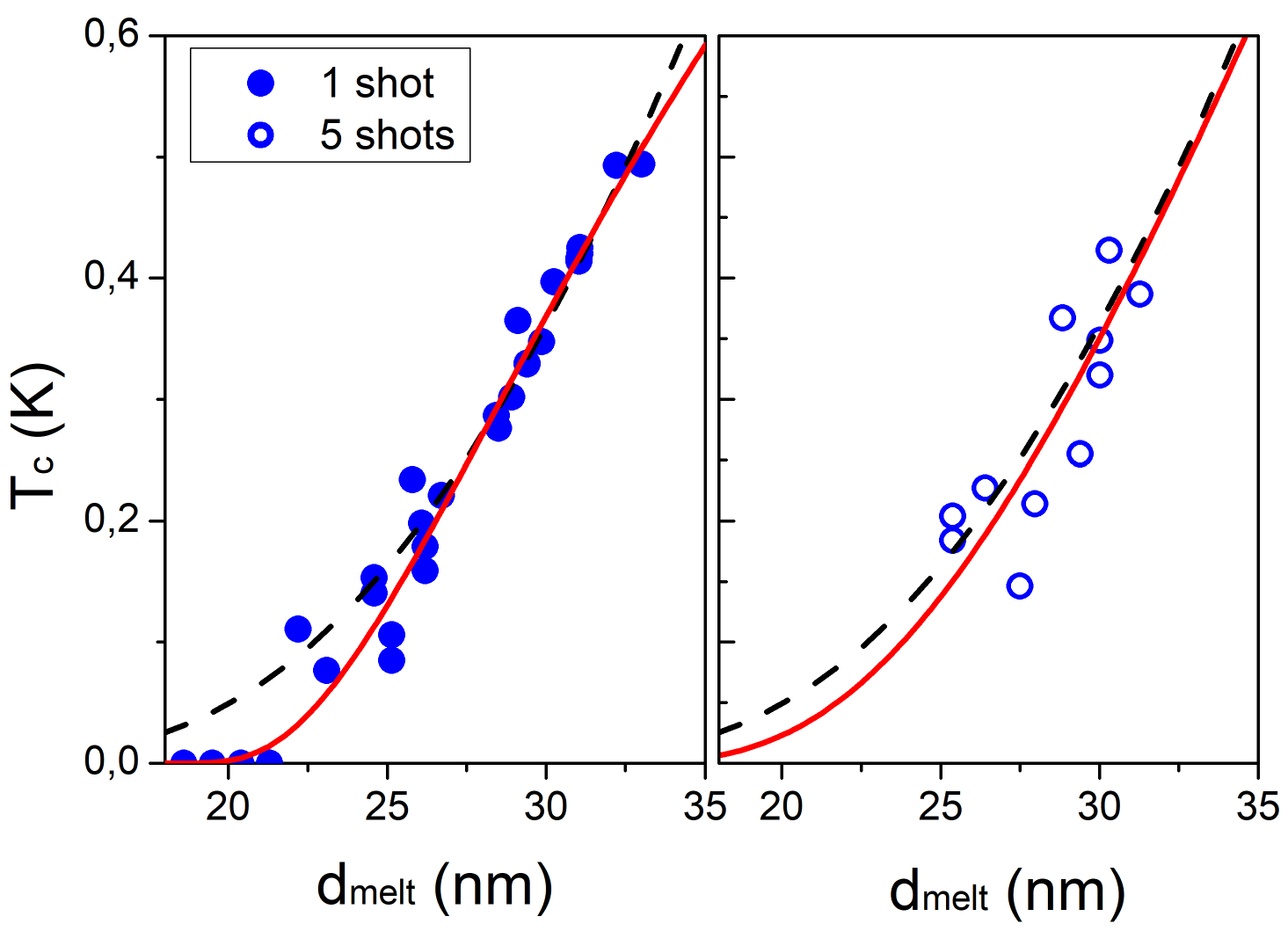}
		\caption{\label{fig:Tc-d-fit} $T_c$ vs laser-melted thickness $d_{melt}$ for both 1 and 5 laser shots. The black dotted line, plotted in both the 1 and 5 laser shots panel, shows the 2-parameter fit, in the hypothesis of $d_S = d_{melt}$, with $T_{c,0}=0.52\,$K and $b=0.5$. The red continuous line shows the 3-parameter fit with the hypothesis that the superconducting layer corresponds to the relaxed portion of the melted layer, with $T_{c,0}=1.04\,$K, $b=0.108$ and $d_S = d_{melt} - 16.6\,$nm for 1 shot and $T_{c,0}=1.04\,$K, $b=0.29$ and $d_S = d_{melt} - 9\,$nm for 5 shots. 
  }
	\end{figure}

We now focus on the strong energy dependence of the superconducting critical temperature. As the laser energy affects directly the melted thickness, we plot $T_c$ vs. $d_{melt}$ in Fig.\ref{fig:Tc-d-fit}. The correspondence between $d_{melt}$ and $E_L$ is extracted independently from the measurements of the TRR, the dose, and the resistance, showing a good agreement and a common dependence $d_{melt} (E_L)$ (details in the Suppl. Mat.).
For both 1 and 5 laser shots, $T_c$ increases steeply with the thickness up to 0.5 K.
To understand the critical temperature dependence with thickness we recall that the melted, ultra-doped layer is on top of the remaining, unmelted Si above the BOX, where an implanted concentration above $\sim 1\,at.\%$ is expected. 
In bilayer structures made of one superconducting (S) layer in contact with a second normal (N) layer, a strong suppression of the $T_c$ can take place, if the interface between the two layers is transparent. This is only observed when the thickness of the S layer is smaller or of the order of the superconducting coherence length, as is our case with $d_{melt} < 33 $nm$ < \xi \approx 50\,$nm \cite{Desvignes2023}. This effect, known as the inverse proximity effect, was already observed on superconducting Si layers realised in bulk Si samples \cite{Grockowiak2013} and is well described with the Usadel model \cite{Fominov2001, Gueron1996, Martinis2000}:
\begin{subequations}
    \begin{align}
         T_{c} & =  T_{c0} \left [\frac{T_{c0} }{1.14\Theta_D}\sqrt{1+\left ( \frac{k_B\Theta_D}{\tau} \right )^2}\right ]^{\frac{b d_N}{d_S}} \\
         \tau & = \frac{\hbar}{2\pi}\frac{v_{F,S}}{\rho_{int}}\frac{b d_N+d_S}{b d_N d_S}\\
         b & = \frac{v_{F,S}}{v_{F,N}} = \left( \frac{n_N}{n_S}\right)^{1/3}
    \end{align}
    \label{eq:Tc_bilayer}
\end{subequations}
where $T_{c0}$ is the superconducting critical temperature of the single S layer, $\Theta_D$ the phonon energy scale in temperature units, $d_N$ ($d_S$)  the thickness, $v_{F,N}$ ($v_{F,S}$) the Fermi velocity and $n_N$ ($n_S$) the active dopant concentration in the normal (superconducting) layer \cite{Chiodi2014}. $\rho_{int}$ is the dimensionless interface resistance per channel, related to the total interface resistance per unit area A by $R_{int} A = h/2e^2 \, (\lambda_{F,S}/2)^2\, \rho_{int}$.\\
The first noteworthy result is that it is possible to describe the $T_c(d_{melt})$ dependence with zero interface resistance ($\rho_{int}=0$), for both 1 and 5 laser shots, highlighting a very good transparency.
Even more, all the fits imposing a small but finite interface are, though in reasonable agreement, less satisfactory than the one at zero interface. Eq.\ref{eq:Tc_bilayer} thus simplifies to: 
\begin{equation}
    T_{c} =  T_{c0}\, \left [\frac{T_{c0}}{1.14\Theta_D} \right ]^{\frac{b d_N}{d_S}}
    \label{TcUsadel}
\end{equation}
To reduce the number of free fitting parameters, we fix $\Theta_D$, of weak influence on the fit, to its simulated value $\Theta_D=650\,$K \cite{Boeri2004}. The simplest thicknesses choice consists in associating the superconducting layer to the melted layer. Thus, the superconducting and normal layer thicknesses are $d_S = d_{melt}$  and $d_N = d_{SOI} - d_{melt}$. The remaining fitting parameters are $T_{c,0}$ and the ratio of the Fermi velocities $b=v_{F,S}/v_{F,N}$. The best fit of $T_c(d_{melt})$ for the whole datasets containing 1 and 5 laser shots results is given by $T_{c,0}=0.52\,$K and $b=0.5$ (Fig.\ref{fig:Tc-d-fit}). The fitted $T_{c,0}$ corresponds well to the critical temperature $T_c = 0.5\,$K measured in GILD Si bulk samples of similar thickness $d=30\,$nm and active doping $n_B=3\times 10^{21}\,$cm$^{-3}$ \cite{Desvignes2023}. Expressing $b$ in in the free electron model $b=v_{F,S}/v_{F,N}=(n_N/n_S)^{1/3}$, we recover a concentration in the unmelted normal layer, underneath the melted layer, of $n_N= 3.75 \times 10^{20} cm^{-3}$. This value is remarkably close to the equilibrium solubility limit $n_{sol}= 4\times 10^{20} cm^{-3}$, suggesting that a large fraction of the dopants present in the implantation queue ($>5.7 \times 10^{20} cm^{-3}$) would be activated, up to the equilibrium saturation concentration, by the heat provided by the melted layer just above (the Si melting temperature being $1683\,$K).  \\
While this two-parameter fit provides a coherent scenario, it is interesting to look at the results of the three parameters fit ($T_{c,0}$, $b$ and $d_S$), relaxing the constraint about the superconducting thickness corresponding to the melted thickness, maintaining $d_N = d_{SOI} - d_S$. The fit is performed on the 1 shot samples, for which a larger range and number of points are available (Fig. \ref{fig:Tc-d-fit}). A better agreement with the experimental data is then achieved, with $T_{c,0}=1.04\,$K, $b=0.108$ and $d_S = d_{melt} - 16.6\,$nm (Fig.\ref{fig:Tc-d-fit}). These fitting parameters suggest a scenario where the superconducting layer is only the top part of the melted layer, above the highest dislocation line, where the strain has been relaxed (Fig.S\ref{fig:Strain}). In this case, the normal layer is $d_N = d_{SOI} - d_{melt} + 16.6\,$nm, which in the case of the samples shown in TEM analysis (Fig.\ref{fig:TEM}) gives $d_N = 20.6\,$nm, in agreement with the position of the dislocation line $17-21\,$nm above the BOX. $T_{c,0}$ is moreover in reasonable agreement to the maximum $T_c = 0.9\,$K observed for $d_{melt} = 300\,$nm thick bulk Si samples, where the influence of the thin, fully strained layer, can be neglected. In order to check if the identification of the superconducting layer with the relaxed one is also coherent in the case of the 5 shots samples, we have plotted in Fig.\ref{fig:Tc-d-fit} the $T_c (d_{melt})$ one parameter fit imposing the same $T_{c,0}=1.04\,$K and $d_N = d_{SOI} - d_{melt} + 9\,$nm, again associating the superconducting layer to the relaxed region (for the sample in Fig.\ref{fig:TEM}, this gives $d_N=13\,$nm in correspondence with the dislocation line $13\,$nm above the BOX). The resulting fitted $b=0.29$ is  then higher than the one estimated from 1 shot samples fit $b=0.108$, in agreement with the larger B concentration expected in the bottom of the sample after 5 laser shots following the more homogeneous distribution of B for multiple laser shots. This results highlight the importance of the structural strain relaxation in the establishment of Si superconductivity, in agreement with recent results on bulk Si \cite{Desvignes2023,Nath2024}.\\ 
\subsection*{Conclusions}
In conclusion, we demonstrate superconducting monocrystalline Si epilayers with critical temperatures up to $0.5\,$K, obtained on 33 nm thick SOI 300 mm wafers after heavy pre-implantation of boron ($2.5\times10^{16} \, cm^{-2})$ followed by nanosecond laser annealing with $25\,$ns pulse duration.
The analysis of the transport properties ($R$, $RRR$, dose $\mathcal{N}$, $T_c$), coupled to structural measurements ($XRD$, $STEM$), have highlighted the effect of the nanosecond laser annealing energy and the impact of multiple laser anneals, both in the normal and superconducting phase.
Increasing the laser energy allows increasing linearly $T_c$, through the increase of the melted thickness (and thus of the superconducting thickness), at a constant saturation active concentration of $6\,at.\%$ (corresponding to the out-of-equilibrium solubility limit attained by nanosecond laser annealing), while also improving the overall quality of the layer, its crystallinity and conductivity. The maximum $T_c$ obtained, $0.5\,K$, is comparable to the $T_c$ of monocrystalline films of similar thickness obtained on bulk Si by Gas Immersion Laser Doping, where the dopants are introduced by 'softer' chemisorption. This implies that nanosecond laser annealing can heal successfully the implantation-induced defects to recover a good crystallinity from the initial amorphized phase. This supports the possibility of transferring the fabrication of nanosecond laser annealed superconducting layers from a laboratory environment towards more standard implantation-based doping techniques.
Performing just a few laser anneals (5) instead of a single one reduces the amount of structural defects, such as stacking faults and dislocations, and homogenizes the B depth distribution.
Finally, the quantitative analysis of the results in the frame of superconductor/normal metal bilayer structures suggests the importance of the structural strain relaxation to achieve the superconducting phase, and demonstrates an excellent transparency between the layers, allowing the further development of superconducting devices on SOI with compatible large scale integration tools. 

%TC:ignore 

\section*{\label{sec:Supplemental} Supplemental materials}

\subsection*{XRD}
\begin{figure}[t]
		\centering
		\includegraphics[width=\columnwidth]{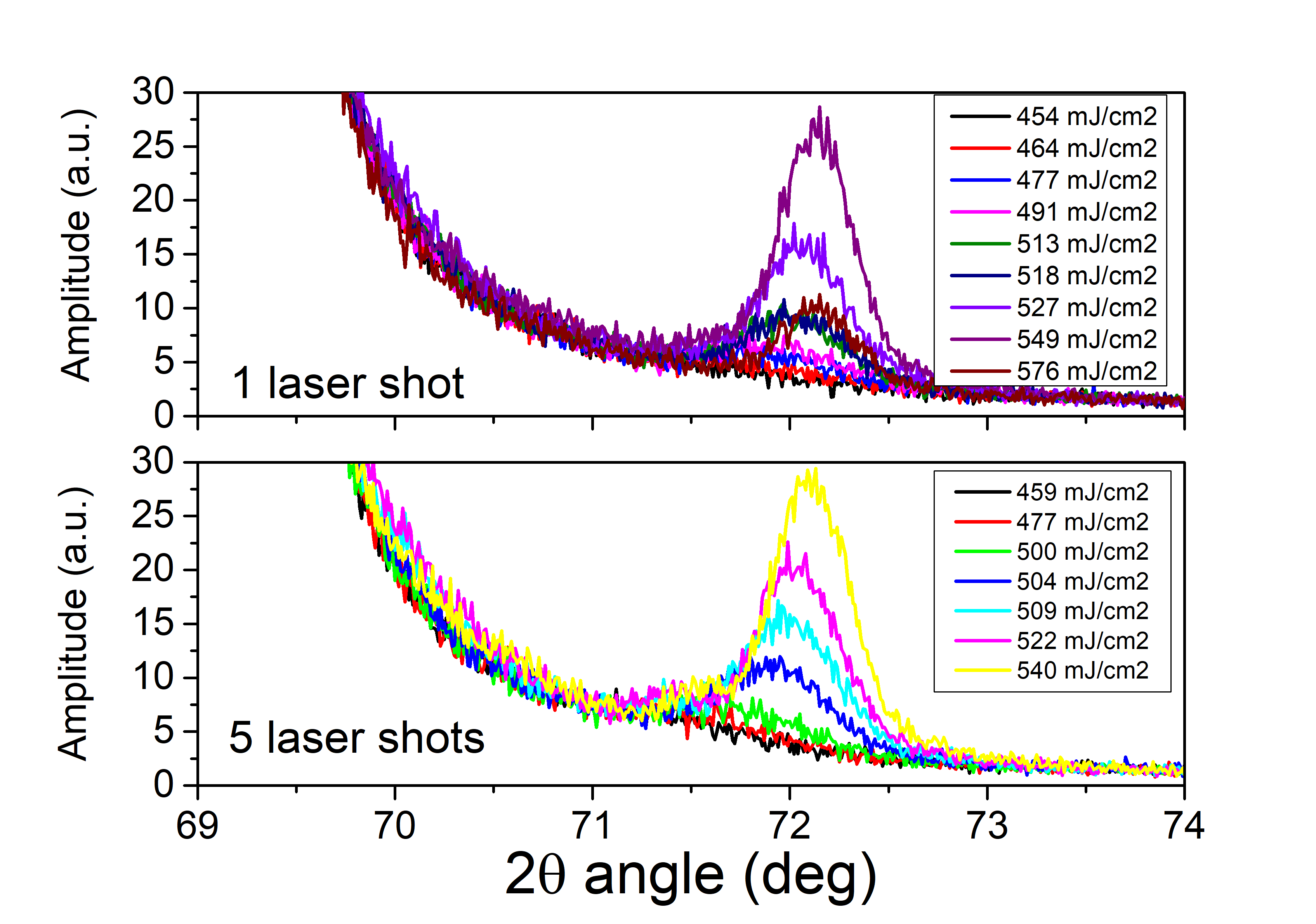}
		\caption{\label{fig:XRD_spectra} Diffraction spectra of laser annealed doped silicon layers. %\textcolor{blue}{add a figure comparing 1 shot and 5 shots at same energy?}
  }
	\end{figure}

\begin{figure}[t]
		\centering
		\includegraphics[width=\columnwidth]{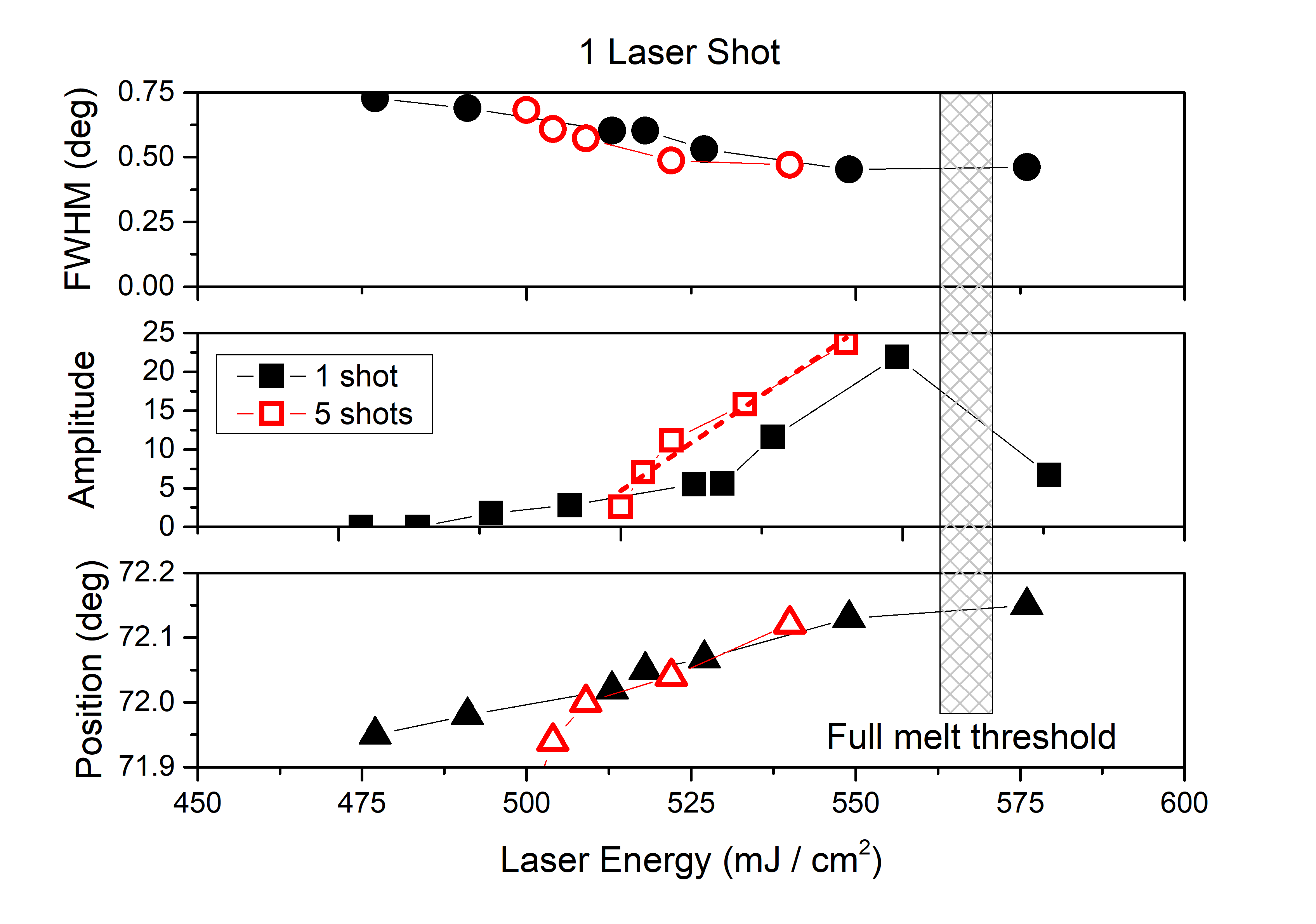}
		\caption{\label{fig:XRD_param} Full width at half maximum (FWHM), amplitude and position of the Si:B XRD diffraction peaks obtained from  Gaussian fits. %\textcolor{blue}{inverse order: position, intensity, FWHM?}
  }
	\end{figure}

We have performed $\theta-2\theta$ X-Rays diffraction scans on 1 and 5 laser shots series. The spectra obtained for 1 and 5 shots are shown in Fig.S\ref{fig:XRD_spectra}. From these curves, the position, width and amplitude have been extracted using a non-linear Gaussian fit in the vicinity of the peaks. The evolution of those quantities is shown in Fig.S\ref{fig:XRD_param}. \\
A clear peak can only be observed when the melt depth exceeds the polycrystalline layer created by explosive recrystallization at low energy. The peak becomes sharper and higher with increasing laser energy (i.e. layer thickness), indicating better crystalline properties resulting in a more homogeneous layer. The position of the Si:B peaks is situated at higher diffraction angle  $2\theta_{SiB}\sim 72^\circ$ than Si, resulting from the reduction of the crystal lattice due to the substitutional incorporation of B atoms.  From the peak position, we have obtained the relative variation of the doped silicon out-of-plane lattice parameter using Bragg law:
\begin{equation}
    \frac{\Delta a}{a} = \frac{a_{perp,SiB}-a_{Si}}{a_{Si}}=(\theta_{SiB}-\theta_{Si})\,cotan(\theta_{Si})
\end{equation}
where $a_{perp,SiB}$ is the Si:B lattice parameter perpendicular to the epilayer, $a_{Si}$ that of the bare silicon and $\theta_{SiB}$ and $\theta_{Si}$ the diffraction angles of Si:B and Si. We obtain $\Delta a/a = -3.4$ to $-3.8 \%$.

\subsection*{Estimating the doped depth}

\begin{figure}[b]
		\centering
		\includegraphics[width=\columnwidth]{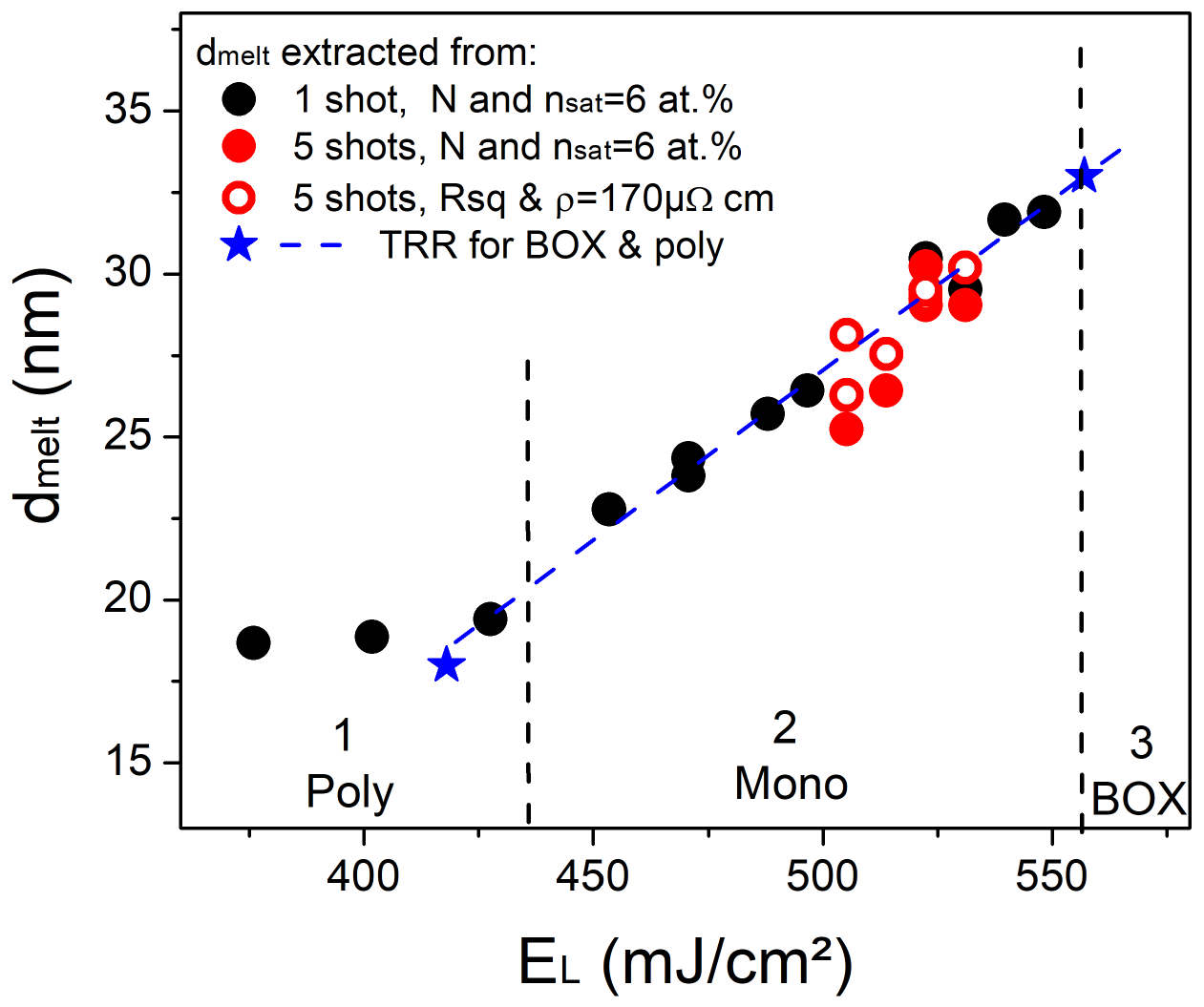}
		\caption{\label{fig:d_Elaser} Estimates of the melted depth using various and independent measurements as a function of the laser energy. Blue stars and blue dotted line: estimation from $TRR$ identification of the melt threshold, poly-Si to c-Si transition and full melt, with $d_{melt} = a E_L +b $, a linear dependence being already observed in \cite{Bonnet2019}. Black and red dots: $d_{melt} = \mathcal{N}/n_{sat}$, with $n_{sat} = 3\times 10^{21}\,at/cm^3$ \cite{Desvignes2023}. Red circles: $d_{melt} = \rho / R_{sq}$ with $\rho = 170 \mu\Omega cm$ \cite{Desvignes2023}. 
  }
	\end{figure}

The estimate of the melted and correspondingly of the recrystallized SiB thickness, is crucial to understand the superconducting properties of our epilayers (see main text). The laser annealed thickness was determined by several means (Fig.S\ref{fig:d_Elaser}). First, the Time Resolved Reflectivity measured in-situ during the doping allows identifying the laser energy $E_L$ for which the melt begins (melting threshold, $d_{melt}=0$), for the poly-Si to monocrystalline Si transition ($d_{poly-mono} \sim d_{a-Si}$) and for the full-melt ($d_{melt} = d_{SOI}$). Since the melted depth is linear with $E_L$ (as confirmed for different laser pulse profiles in \cite{Bonnet2019}), a first estimate of $d_{melt} (E_L)$ is realized (stars and blue dotted line, Fig.S\ref{fig:d_Elaser}). This first study also shows that both the thinnest ($d_{poly-mono}$) and thickest ($d_{full-melt}$) monocrystalline layers display a similar concentration $\mathcal{N}/d \approx 6 at.\% = 3\times 10^{21}\,at/cm^3$. Assuming that all the layers have attained the same limit saturation concentration, it is possible to plot the melted depth $d_{melt} = \mathcal{N}/n_{sat}$ from the whole set of active dose $\mathcal{N}$ measurements, for both 1 and 5 laser shots (Fig.\ref{fig:Rs_Tc_Dose_vs_Elaser}), confirming the $TRR$ estimation  (black and red dots, Fig.S\ref{fig:d_Elaser}).  The constant active boron concentration is also in agreement with XRD measurements. \\   
Additionally, the melted depth can also be extracted from the square conductance : $d_{melt}=\rho G_{sq}$ with $\rho$ the resistivity of the recrystallized SiB layer assuming the underneath layer has a negligible conductance (red circles, Fig.S\ref{fig:d_Elaser}). For this extraction we only use data from the homogeneous 5 laser shots series. The value of the resistivity, $\rho = 170 \,\mu \Omega cm$, is fixed from the resistivity of the full melt sample, for which the thickness is known exactly, and whose value is coherent with previous studies \cite{Desvignes2023}.\\
Remarkably, these estimates of the melted depth extracted from different measurements are very consistent with each others and within 10$\%$ uncertainty. Moreover, the melted depth dependence is consistent with the process described in section\ref{regimes}: at low energy, $d_{melt}$ is almost constant and close to $18 \, nm$, while at larger energy, it increases linearly to reach $33\,nm$ at the full melt threshold.

\subsection*{TEM - structural deformation}
\begin{figure}[t]
		\centering
		\includegraphics[width=\columnwidth]{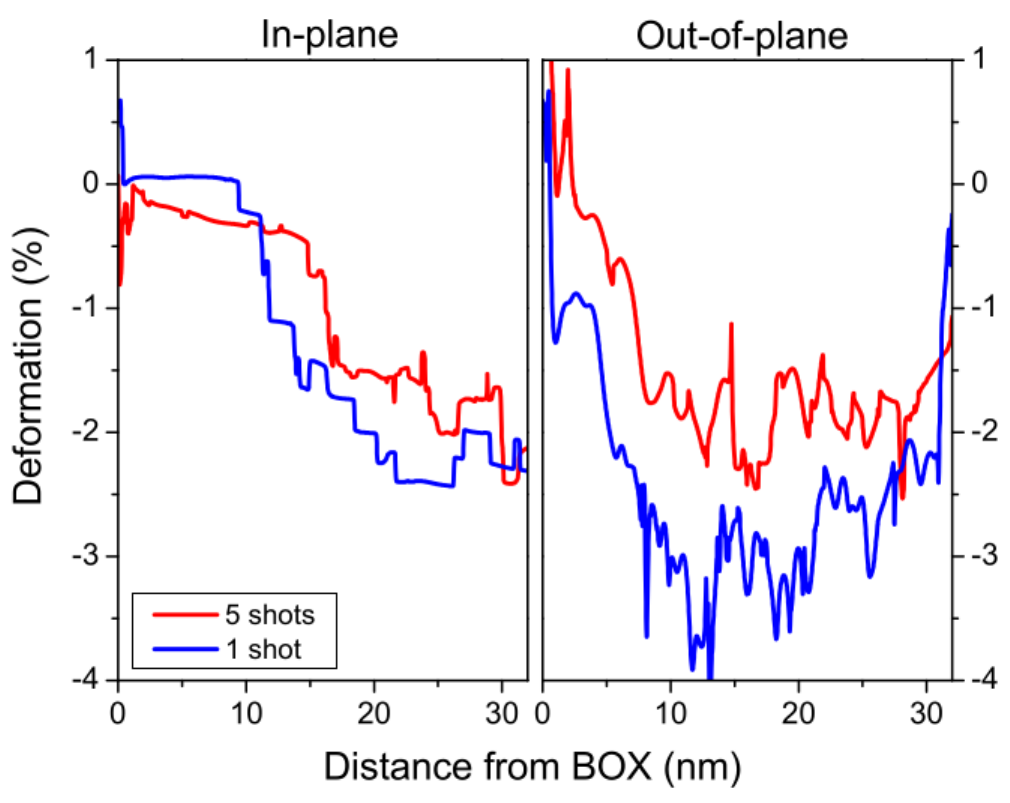}
		\caption{\label{fig:Strain} In plane and out-of-plane lattice deformation, obtained by GPA method from HRTEM images \cite{Hytch1998}, as a function of the distance to the BOX, for the two samples shown in Fig.\ref{fig:TEM} with 1 or 5 laser shots at $E_L=522 mJ/cm^2$ ($d_{melt}$=29 nm). }
	\end{figure}

The in-plane and out-of-plane lattice deformations were extracted by the Geometrical Phase Analysis (GPA) method from High Resolution Transmission Electron Microscope (HRTEM) images. Fig.S\ref{fig:Strain} shows the deformation for the two samples shown in Fig.\ref{fig:TEM}. 
The higher density of structural disorder can be correlated to the larger lattice deformation present in 1 shot samples. Indeed, the in-plane and out-of-plane deformations $\epsilon_{//}=(a_{SiB//}-a_{Si})/a_{Si}$ and $\epsilon_{\perp}=(a_{SiB\perp}-a_{Si})/a_{Si}$  measured on the samples of Fig.\ref{fig:TEM}, show $\epsilon_{//}=$1.9$\%$ (1.5$\%$) and $\epsilon_{\perp}=$2.5$\%$ (2$\%$) for 1 shot (5 shots) respectively.

\bibliographystyle{unsrt}
\bibliography{biblio}% Produces the bibliography via BibTeX.
%TC:endignore 
\end{document}